\documentclass[12pt]{iopart}

\usepackage{iopams}  
\usepackage{graphicx}

\def\laq{~\raise 0.4ex\hbox{$<$}\kern -0.8em\lower 0.62ex\hbox{$\sim$}~}
\def\gaq{~\raise 0.4ex\hbox{$>$}\kern -0.7em\lower 0.62ex\hbox{$\sim$}~}

\def\beq{\begin{equation}}
\def\eeq{\end{equation}}
\def\bea{\begin{eqnarray}}
\def\eea{\end{eqnarray}}
\def\bean{\begin{eqnarray*}}
\def\eean{\end{eqnarray*}}

\def \pa {\partial}

\def \La {\Lambda}
\def \Da {\Delta}

\def \ga {\gamma}

\def \r {\rho}

\def \Om {\Omega}

\newcommand{\bb}{\bibitem}

\begin{document}

\begin{flushright}
BA-TH/672-13\\
To appear in {\bf Class. Quantum Grav.}
\end{flushright}

%\vspace{-0.5cm}

\title{Discriminating different models of luminosity-redshift distribution}

\author{L. Cosmai, G. Fanizza, M. Gasperini and L. Tedesco }

\address{Dipartimento di Fisica, Universit\`a di Bari, Via G. Amendola
173, 70126 Bari, Italy,\\
and Istituto Nazionale di Fisica Nucleare, Sezione di Bari, Bari, Italy}

\ead{leonardo.cosmai@ba.infn.it, giuseppe.fanizza@ba.infn.it, gasperini@ba.infn.it, Luigi.Tedesco@ba.infn.it}

\begin{abstract}

The beginning of the cosmological phase bearing the direct kinematic imprints of supernovae dimming may significantly vary within different models of late-time cosmology, even if such models are able to fit present SNe data at a comparable level of statistical accuracy. This effect -- useful in principle to discriminate among different physical interpretations of the luminosity-redshift relation -- is illustrated here with a pedagogical example based on the LTB geometry. 

\end{abstract}

%\pacs{98.80.-k, 98.80.Es, 95.36.+x}
%\submitto{\CQG} 
%\maketitle

%\section{Introduction}
%\label{sec1}

%\vskip 1 cm
It is by now widely known that the observed luminosity-redshift distribution of type Ia supernovae \cite{1} can be fitted even without dark energy, provided one introduces a sufficiently inhomogenous space-time geometry. A typical, very simple example of such a possibility is provided by matter-dominated cosmological models of the  Lema\^itre-Tolman-Bondi (LTB) type (see e.g. \cite{2} for an incomplete list of papers on this subject), provided the observer is located near enough to the symmetry centre of the inhomogeneous -- but isotropic -- matter distribution \cite{3,3a}. 

Such an example may be regarded as unnatural because of the amount of fine tuning required to localize the observer position \cite{3a}, and also appears theoretically disfavoured by the possible presence of weak geometric singularities \cite{3b}. Nevertheless, the possible role of inhomogeneities in determining (or at least substantially contributing to)
the large-scale dynamics should be -- and, indeed, currently is \cite{3c} -- seriously scrutinized and discussed, even in the presence of a dominant dark-energy cosmic component \cite{3d}.

The general question that arises in such a context is how to distinguish different successful analyses of the SNe data based on different physical models and, in particular, on different (homogeneous versus inhomogeneous) large scale geometries. For inhomogeneous models of the LTB type various answers to this question are known, concerning the size of the voids described by the LTB geometry \cite{4}, the local expansion rate inside the voids \cite{5}, and the associated effect of redshift drift \cite{6}. Other possibilities to test LTB models are provided by studies of scalar perturbations \cite{7}, of small-scale CMB effects \cite{8}, of the cosmic age parameter \cite{9}, and of BAO (baryon acoustic oscillations) data \cite{10}.

The main purpose of this note is to point out another possible difference between inhomogeneous and more conventional interpretations of the SNe data, not yet discussed in the literature; such a difference is based on the value of the redshift parameter $z_{\rm acc}$ (to be defined below, see after Eq. (\ref{14})), marking the beginning of the regime directly characterized by the kinematic imprints of SNe dimming. The value of such parameter can be largely different even within models able to fit the presently observed luminosity-redshift distributions at a comparable level of accuracy (see e.g. \cite{11} for earlier studies on the beginning of the accelerated regime in the context of a homogeneous geometry). 

This suggests two possible experimental ways of discriminating among  models of the luminosity-redshift relation. First, direct  observations able to extend our present knowledge of the Hubble diagram up to values of $z$  higher than those allowed by present SNe data: for instance, gamma-ray burst (as discussed in \cite{12}), or even gravitational waves observations, through an analysis of the luminosity distance of the so-called ``standard sirens'' \cite{13}. Second, indirect observations which are sensitive to the time-dependence of the so-called ``transfer function" \cite{13a}, which controls the evolution of the primordial perturbation spectrum inside the horizon down to the present epoch, and which is crucially affected by the kinematics of the cosmological background (see e.g. \cite{13b}). 

The possible relevance of the parameter $z_{\rm acc}$ will be illustrated in this paper by a simple exercise,  in which the SNe data of the recent  Union2 compilation \cite{14} are fitted using a inhomogeneous, matter-dominated LTB model, and such a fit is compared  with the standard one performed in the context of the flat concordance $\La CDM$ model. 
We stress that our aim {\em is not} to provide a realistic alternative to the successful concordance cosmology, but only to discuss how to distinguish, at least in principle, different fits of  SNe data based on different geometric schemes.  The proposed diagnostic may be added to 
other general methods aiming at discriminating the expansion history of competing models,  like -- in particular --  dark-energy based diagnostics for homogeneous models \cite{20a}; a Friedmann equation diagnostic for homogeneous versus inhomogeneous models \cite{20b,20c}; and the already mentioned test of redshift drift \cite{6,20c,20d}. 

The  cosmological configuration we will consider refers to a late-time (in particular, post-reionization) Universe, characterized by a stochastic distribution of many overdense and underdense regions, of various possible sizes and shapes, possibly even incoherently superimposed among each other\footnote[7]{Such a configuration is in principle different from that of a typical ``Swiss cheese'' scenario, where the void regions are more or less regularly distributed and well disconnected (see e.g. \cite{15}).}. Let us suppose that in such a context, and up to a given scale $r_V$ (to be specified below), the effective (averaged) large-scale geometry can be {\em locally} described by a model of the LTB type. Such a model is characterized in general by three arbitrary functions of the radial coordinate (see e.g. \cite{15a}). For the illustrative purpose of this paper, however, it will be enough to consider a simple example where the contribution of the spatial curvature is negligible and the gravitational sources are dominated by an isotropic cold dark matter (CDM) distribution (but the model could be easily generalized by the addition of an arbitrary cosmological constant). 

We will assume that the large-scale geometry around a given observer is described -- in polar coordinates and in the synchronous gauge --  by the following metric, 
\beq
ds^2= dt^2- A^{\prime 2} (r,t) dr^2 - A^2(r,t) \left( d \theta^2 + \sin^2 \theta d\phi^2\right),
\label{1}
\eeq
where a prime denotes partial derivatives with respect to $r$ and a dot  with respect to $t$. In the limit $A(r,t)= r a(t)$ one recovers the well known, spatially flat, Friedman-Lema\^itre-Robertson-Walker (FLRW) metric. In general, the unknown function $A(r,t)$ is to be determined by the Einstein equations, which in our case reduce to
\beq
H^2 +2 HF = 8 \pi G \r, ~~~~~~~~~~~~~~~
2 \dot H+3 H^2=0,
\label{2}
\eeq
where
$H(r,t)=\dot A / A$ and $F(r,t)=\dot A' /A'$. The density profile of the CDM distribution around a central observer,  $\r =\r(r,t)$ satisfies the covariant  conservation equation:
\beq
\dot \r+\left(2H+F \right)\r =0,
\label{3}
\eeq
while all the other Einstein equations are identically satisfied by the metric (\ref{1}) (see e.g. \cite{16}).

The above cosmological equations can be integrated exactly, and in this paper we will adopt the particular exact solution 
\beq
A(r,t)= r \left[1+{3\over 2} tH_0(r) \right]^{2/3},
\label{4}
\eeq
normalized in such a way that $A=r$ at $t=0$. The arbitrary function $H_0(r)$ depends only on the radial coordinate, and the usual matter-dominated FLRW solution is exactly recovered in the limit  $H_0=$ const. We will use, in particular, the parametrization
\beq
H_0(r)= \overline{H} + \Da H e^{-r/r_V},
\label{5}
\eeq
already suggested in \cite{16} for a similar LTB scenario (a brief discussion of other possible choices for the phenomenological profile $H_0(r)$ will be given in the final part of this paper). For the chosen profile the combination of parameters $\overline{H} + \Delta H \equiv H_0(0)$ corresponds to the locally measured value of the Hubble constant, while the distance  $r_V$  represents the typical distance scale above which inhomogeneity effects become rapidly negligible.  

To make contact with more general forms of the LTB metric appearing in the literature, and expressed in terms of three functions $M(r)$, $t_B(r)$, $E(r)$, (we are following the notations of \cite{15a}), it may be useful to report here the values of those functions for the model we are using. The effective gravitational mass with comoving radius $r$, for our solution, is given by $M(r)=(1/2) r^3 H_0^2(r)$. It can be easily checked that this function grow like $r^3$ for $r \ll r_V$ and $r \gg r_V$, while, in the transition regime $r \sim r_V$, it is characterized by a fractal index $D=0.4$, i.e. $M(r) \sim r^{3-D} \sim r^{2.6}$. The time scale $t_B$ -- i.e., the local ``big-bang time" at which $A(r,t)=0$ -- in our case is given by $t_B(r)=-(2/3) H_0^{-1}(r)$.

Finally, it is important to stress that the obtained solution is consistent with our assumption of vanishing spatial curvature, i.e. with the choice $E(r)=0$. Perturbing the solution with the addition of scalar curvature (and assuming that $E(r) \sim r^2$ as in the large-scale FLRW limit), we have checked indeed that the curvature contribution to the total energy density may have a variation 
which is at most of the order of $0.05 \%$ over length scales of order $r_V$ and time scales of order $H_0^{-1}$. Hence, if initially small but nonzero, it keeps small over the whole spatial and temporal range of interest for this paper. 

Let us now compute the luminosity distance $d_L$ of a source emitting light at a cosmic time $t$ and a radial distance $r$ from the origin. We will assume, for the moment, that the observer is also located at the origin (the consequences of a possible off-center position will be discussed later). The angular distance (or area distance) of the source, for the metric (\ref{1}), is then given by $d_A= A(r,t)$, and the luminosity-distance, according to the so-called ``reciprocity law" \cite{18}, reduces to $d_L= (1+z)^2 A(r,t)$, where $z$ is the redshift parameter evaluated along a null radial geodesic connecting the source to the origin. 

Calling $u^\mu$ the static (time-like) geodesic vector field tangent to the worlines of  source and observer, and $k^\mu$ the null vector tangent to the null radial geodesic, we find in our metric $u^\mu= dx^\mu/ d \tau= (1, 0, 0,0)$ and $k^\mu= \left( (A')^{-1}, -(A')^{-2}, 0,0\right)$. Hence, for light emitted at time $t$, radial position $r$, and observed at the origin at $t=t_0$,
\beq
1+z=  {\left(k^\mu u_\mu \right)_{r,t}\over \left(k^\mu u_\mu \right)_{0,t_0}}= {A_0'\over A'(r,t)},
\label{6}
\eeq
where $A'_0 \equiv A'(0,t_0)=$ const.

For the phenomenological applications of this paper we need to express $d_L$ completely in terms of the redshift, namely we need to invert Eq. (\ref{6}) to determine $r(z)$ and $t(z)$. We may consider, to this purpose, the differential variation of $z$ with respect to the proper time interval $d \tau$ separating two different instants of light emission, at fixed observation coordinates: $dz/d \tau= u^\mu \pa_\mu z= -(1+z) \dot A'/A'$. It follows that, along a null radial geodesic (where $dt=- A'dr$):
\beq
{dt\over dz}= {dt\over d\tau}{d\tau\over dz}= -{A'\over (1+z) \dot A'},
~~~~~~~~~~
{dr\over dz}= -{1\over A'} {dt \over dz} =-{1\over (1+z) \dot A'}.
\label{7}
\eeq
For the model of Eq. (\ref{4}), in particular, we obtain the differential equations
\bea
\frac{dt}{dz}=& -\frac{1}{1+z}\,\frac{[2+3tH_0(r)][2+3 \,  t \, H_0(r)+2rtH_0'(r)]}{6t H_0^2(r)+4 rH_0'(r)+4 H_0(r)[1+rtH_0'(r)]},
\nonumber \\
\frac{dr}{dz}=& \frac{1}{2^{1/3}(1+z)}\, \frac{[2+3t H_0(r)]^{4/3}}{3 t H_0^2(r)+2rH_0'(r)+2H_0(r)[1+r \, t H_0'(r)]}.
\label{8}
\eea
Solving the above equations for $t(z)$, $r(z)$, and inserting the solutions into the explicit definition of $d_L$, 
\beq
d_L(z)= (1+z)^2A(r,t) = (1+z)^2\, r(z) \left[1+ {3\over 2} t(z) H_0(r(z)) \right]^{2/3}, 
\label{9}
\eeq
we are now in the position of comparing the predictions of our model with the observational data (as well as with the predictions of the standard $\La$CDM scenario). 

Let us first recall that the Union2 compilation of the Supernova Cosmology Project \cite{14} concerns redshift-magnitude measurements of $557$ SNe of type Ia and provides, for each supernova, the observed distance modulus (with relative error) $\mu^{\rm obs} (z_i) \pm \Da \mu (z_i)$, $i=1, \dots , 557$, for redshift values ranging from $z_1=0.015$ to $z_{557}=1.4$. The distance modulus $\mu(z)$ controls the difference between apparent and absolute magnitude, and is related to the luminosity distance $d_L(z)$ by:
\beq
\label{10}
\mu(z) = 5 \, \mathrm{{log_{10}}} \left[\frac {d_L(z)}  {1\,\mathrm{Mpc}}  \right] + 25.
\eeq
Here $d_L$ is given in units of Mpc, and the constant number $25$ is determined by the conventional reference scale assumed for the absolute magnitude.

\begin{figure}[h!]
\centering
\vspace{0.3 cm}
\includegraphics[height=4.5 cm]{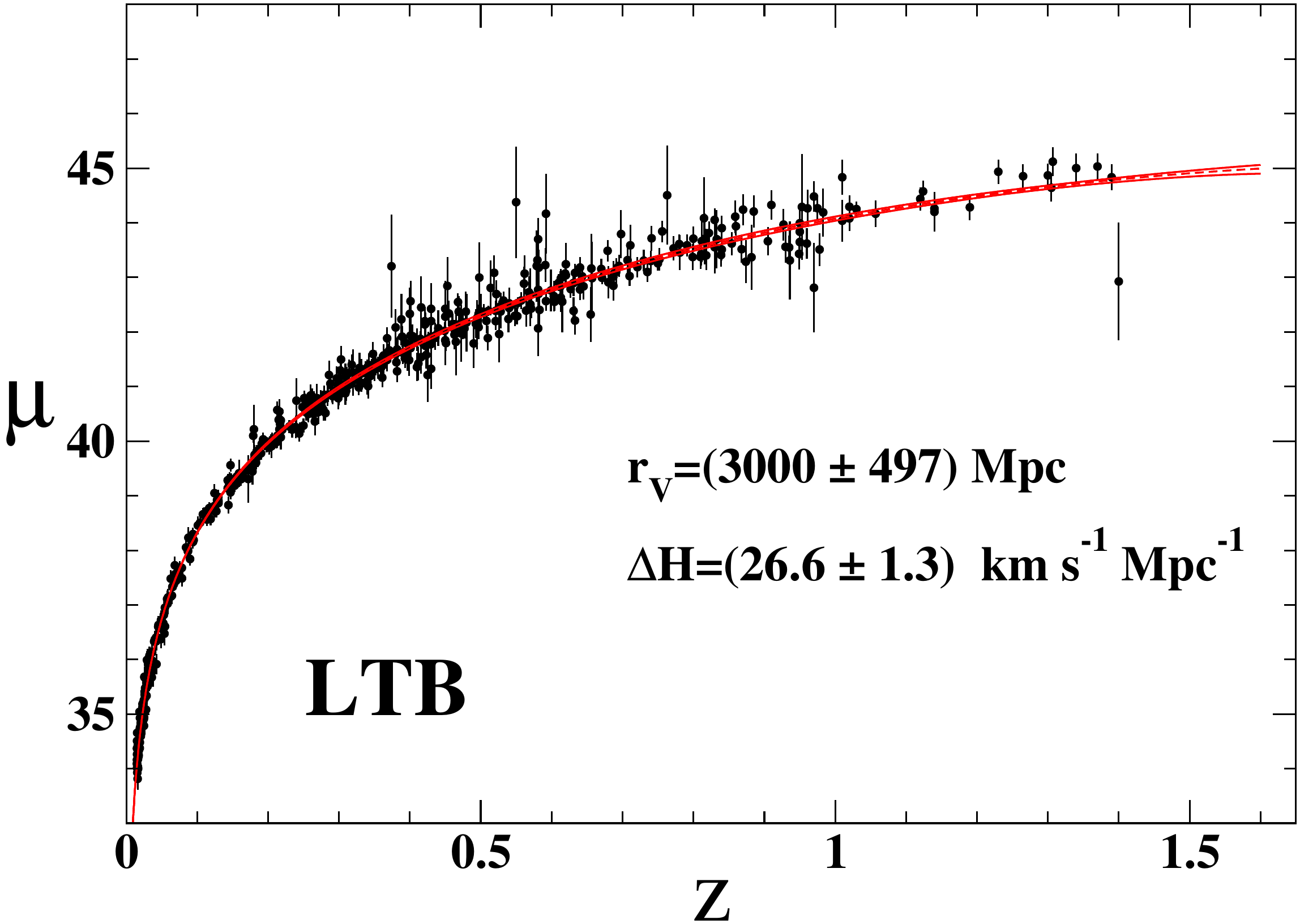}~~~~
\includegraphics[height=4.5 cm]{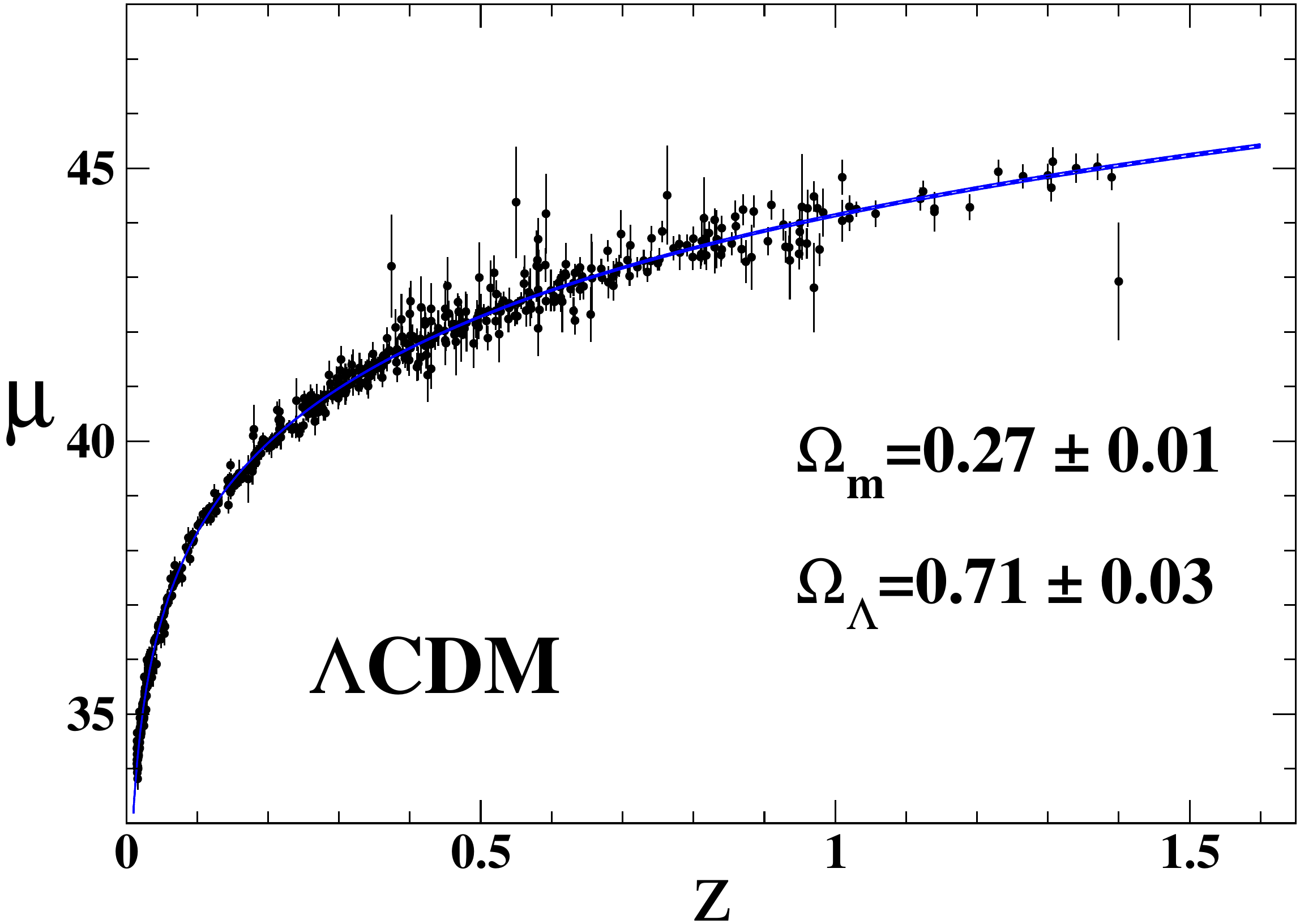}
\caption{The Hubble diagram of the Union2 dataset. The left panel illustrates the  best-fit result for a two-parameter fit of our example of inhomogeneous geometry, with $\chi^2_{LTB}/{\rm d.o.f.}=0.99$. In the right panel we present the corresponding best-fit result for a homogeneous $\La$CDM model, with  $\chi^2_{\La{\rm CDM}}/{\rm d.o.f.}=0.98$.}
\label{Fig1}
\end{figure}

The luminosity distance of Eq. (\ref{9}), with $H_0(r)$ given by Eq. (\ref{5}), is characterized in principle by three independent parameters, and can be applied to fit the experimental data by allowing free variations of $\overline H$, $\Da H$ and $r_V$.  We have performed that exercise, and  found  that  the resulting best fit provides for $H_0(0) \equiv \overline{H} + \Da H$ a value very close to $70 \,{\rm Km \, s}^{-1} {\rm Mpc}^{-1}$. We have thus chosen to concentrate the present discussion on a simpler, two-parameter fit of the data -- which, in any case, is sufficiently accurate for the illustrative purpose of this paper -- by imposing on our model the ``a priori" constraint $\overline{H} + \Da H= 
70 \,{\rm Km \, s}^{-1} {\rm Mpc}^{-1}$. In this way we can eliminate, for instance, $\overline H$, and we can fit the experimental points   $\mu^{\rm obs} (z_i) \pm \Da \mu (z_i)$ by performing a standard $\chi^2$ analysis with
 \beq
\label{11}
\chi^2 = \sum_{i=1}^{557} \left[ \frac {\mu^{\rm obs} (z_i) - \mu(z_i, r_V, \Delta H)} {\Da \mu(z_i)}\right]^2 \, .
\eeq
The theoretical values $ \mu(z_i, r_V, \Delta H)$ can be determined, for each value of $z_i$, by numerically integrating the two equations  (\ref{8}), and computing the corresponding  $d_L(z_i)$ as a function of the two parameters $r_V, \Da H$. By minimizing the above $\chi^2$ expression we have found the best fit values
\beq
r_V= 3000 \pm 497\, {\rm Mpc}, ~~~~~~~~~~
\Da H= 26.6 \pm 1.3\, {\rm Km \, s}^{-1} {\rm Mpc}^{-1},
\label{12}
\eeq
at a confidence level of $95 \%$, and with a goodness of fit $\chi^2/{\rm d.o.f.}=0.99$. The minimization has been performed using the MINUIT package from CERNLIB \cite{18a}. The result of the fit is graphically illustrated by the red curve plotted in the left panel of Fig. \ref{Fig1},  superimposed to the full set of Union2 data (reported with error bars).

Consider now, for comparison, a fit of the same data performed in the context of a spatially flat FLRW geometry, with perfect fluid sources representing CDM and a cosmological constant $\La$. Denoting with $\Om_m$ and $\Om_\La$ the present critical fraction of dark matter and dark energy, we can express the luminosity distance in the usual integral form as 
\beq
d_L(z)= {1+z\over H_0} \int_0^z dx \left[\Om_m (1+x)^3 + \Om_\La \right]^{-1/2}
\label{13}
\eeq
(see e.g. \cite{19}). 
Proceeding as in the previous case, we will reduce the number of parameters from $3$ to $2$ by imposing the same phenomenological constraint as before, which in this case amounts to the condition $H_0 (\Om_m + \Om_\La)^{1/2}= 70 \,{\rm Km \, s}^{-1} {\rm Mpc}^{-1}$. Using Eq. (\ref{13}) to compute $\mu(z_i, \Om_m, \Om_\La)$, and minimizing the corresponding $\chi^2$ expression, we obtain the best fit values $\Om_m=0.27 \pm 0.01$, $\Om_\La=0.71 \pm 0.03$, at a confidence level of $95 \%$, with $\chi^2/{\rm d.o.f.}= 0.98$. The result of the fit is illustrated by the blue curve on the right panel of Fig. \ref{Fig1}. 

%%%%%%%%%%%%%%%%%%%

%%%%%%%%%%%%%%%%%%%%%%%%%%%%

The luminosity-redshift relations of the two models of Fig. \ref{Fig1} are in good agreement with the data, and in both cases the data points are fitted at a comparable level of statistical accuracy. However, we can disclose an important physical difference between the two fits if we subtract from the distance modulus of the two models the distance modulus $\mu^{\rm Milne}(z)$ of a linearly expanding (but globally flat) homogeneous Milne geometry (see e.g \cite{20}), namely if we consider the quantity 
\beq
\Da(z)= \mu(z)- \mu^{\rm Milne}(z)= 5 \, \mathrm{{log_{10}}} \left[\frac {d_L(z)}  {1\,\mathrm{Mpc}}  \right] - 
5 \, \mathrm{{log_{10}}} \left[\frac {z(2+z)}  {2 H_0 \,\mathrm{Mpc}}  \right] ,
\label{14}
\eeq
where $H_0$ is given in units of Mpc$^{-1}$. It is clear that positive or negative values of $\Da$ correspond to luminosity distances which are -- at a given fixed $z$ -- respectively larger or smaller than the reference values of the Milne model.

The case $\Da <0$ is typical of a decelerated Universe like that described by the standard cosmological scenario, where, at the same fixed $z$, the distances are {\em smaller} (or the received fluxes of radiation, i.e. the apparent magnitudes, are {\em larger}) than predicted by a linearly expanding model. The case $\Da>0$, on the contrary, corresponds at the same $z$ to {\em larger} distances (or {\em smaller} radiation fluxes) than predicted by  linear expansion, and is only possible if the model undergoes a period of ``effective" accelerated expansion. In this last case, the transition across the value $\Da=0$ defines an epoch -- characterized by the parameter $z_{\rm acc}$ such that $ \Da (z_{\rm acc})=0$ -- marking the beginning of the cosmological phase directly imprinted by the kinematic effects of the acceleration.

\begin{figure}[h!]
\centering
\includegraphics[height=5.7 cm]{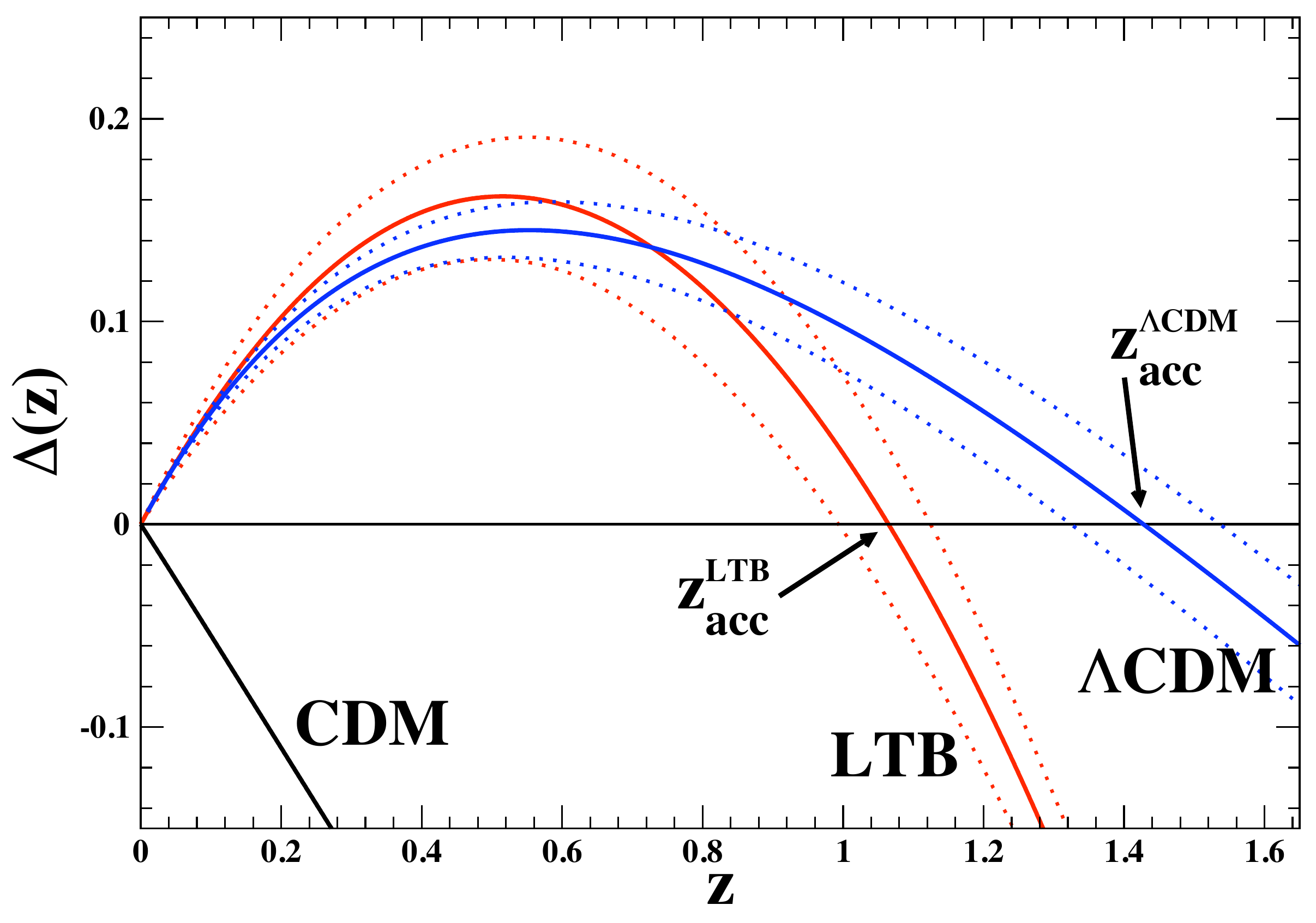}
\caption{The parameter $\Da(z)$ of Eq. (\ref{14}) for the two best-fit models of Fig. \ref{Fig1}. In both cases we have shown the region allowed by the fit at the $95 \%$ C.L. (bounded by dotted lines). We have also reported (for comparison, and without error band) the case of the standard CDM model with $\Om_m=1$.}
\label{Fig2}
\end{figure}

The plot of $\Da(z)$ is presented in Fig. \ref{Fig2} for three cases: the standard CDM-dominated (always decelerated) model, and the two best-fit models of Fig. \ref{Fig1} (corresponding to our  example of inhomogeneous geometry and to a typical example of homogeneous concordance cosmology). 
In the last two cases we have plotted the central values of the fit (solid curves), as well as  the corresponding error bands\footnote[7]{The error bands have been numerically computed  by varying the fit parameters  within the range determined by the corresponding estimated errors.  
All possible fitting curves lying inside the given error region satisfy 
the constraint: $\chi^2  \le \widetilde\chi^2  + \delta$, where $\widetilde\chi^2$ is the value obtained by minimizing Eq.(11) and the constant $\delta$ depends
on the number of parameters and on the confidence level of the fit determined by the MINUIT package \cite{18a}. In our case, in particular, $\delta = 5.99$ for a confidence level of $95\%$.} at the $95 \%$ level of confidence (bounded by the dotted curves). 

%%%%%%%%%%%%%%%%%%%%%%%%%

%%%%%%%%%%%%%%%%%%%%%%%%%

We can see from Fig. \ref{Fig2} that $\Da(z)$ is always negative for  the CDM model, as expected. For the other two models, instead, we have $\Da(z)>0$ in the redshift range $z<z_{\rm acc}$ (because, as expected, a successful fit of the SNe data requires the presence of a phase describing -- or mimicking -- accelerated expansion). However, the values of $z_{\rm acc}$ defined by the condition $\Da (z_{\rm acc})=0$ are largely different in the two models. We find, in particular,
\beq
z_{\rm acc}^{\rm LTB}=1.07 \pm 0.06, ~~~~~~~~~~~~~~~~~~~~
z_{\rm acc}^{\La{\rm CDM}} =1.43 \pm 0.10,
\label{15}
\eeq
and this difference falls outside the error bands illustrated in Fig. \ref{Fig2}
(it is also much larger than the experimental uncertainty affecting present redshift measurements). This suggests that a precise (near-future?) determination of this parameter could provide a clear physical discrimination among different models implementing successful (and statistically equivalent) fits of SNe data. 

It should be mentioned, at this point, that in the computations of the error bands we have neglected the dispersion of data due to the possible presence of a cosmic background of stochastic perturbations: indeed, such a background may induce large errors at very small $z$, but in the range $z \sim 1$ (typical of $z_{\rm acc}$) the induced errors are typically lying in the few-percent range \cite{3d}, hence are not expected to have a crucial impact on the results illustrated in Fig. \ref{Fig2}. The same is expected to be true for the systematic errors -- possibly slightly bigger than the previous ones, but in any case $\laq 10 \%$ -- induced on $z_{\rm acc}^{\rm LTB}$ (but not on $z_{\rm acc}^{\La{\rm CDM}}$) by methods of SNe data reduction based on the assumption of standard homogeneous cosmology (and used in particular for the Union2 catalogue, see e.g. \cite{21a}). Finally, we should note that a value of $z_{\rm acc}$ compatible with that of the inhomogeneous model considered here could be reproduced also in a homogeneous $\La$CDM context, with realistic values of $\Om_m$ and $\Om_\La$, but only at the price of introducing a large enough {\em negative} spatial curvature, with $\Om_k \sim 0.1$ (for instance, a model with $\Om_m=0.3$, $\Om_\La=0.6$, $\Om_k=0.1$ gives $z_{\rm acc}^{\La{\rm CDM}}=1.087$).

\begin{figure}[h!]
\centering
\includegraphics[height=5.7 cm]{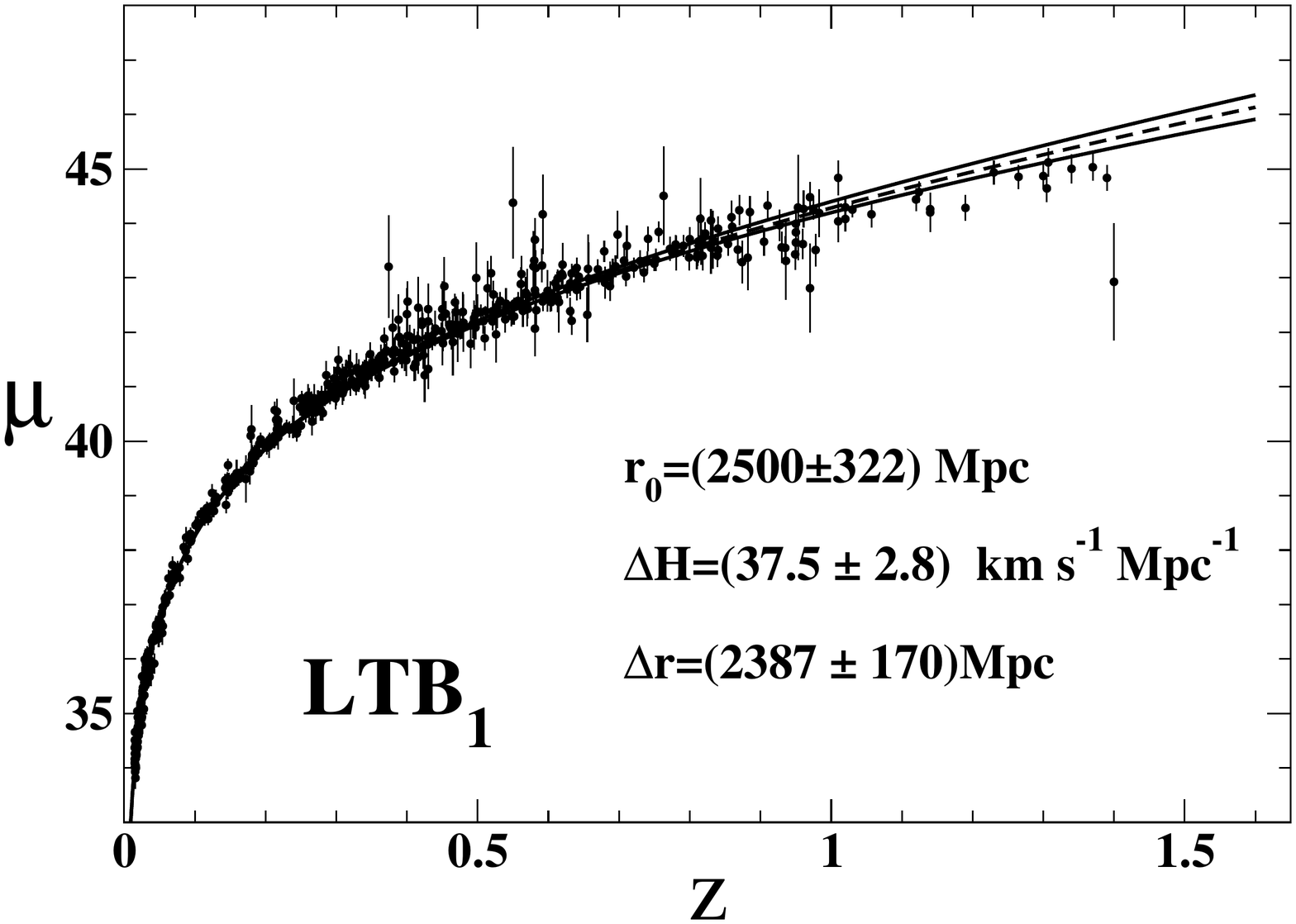}
\includegraphics[height=5.7 cm]{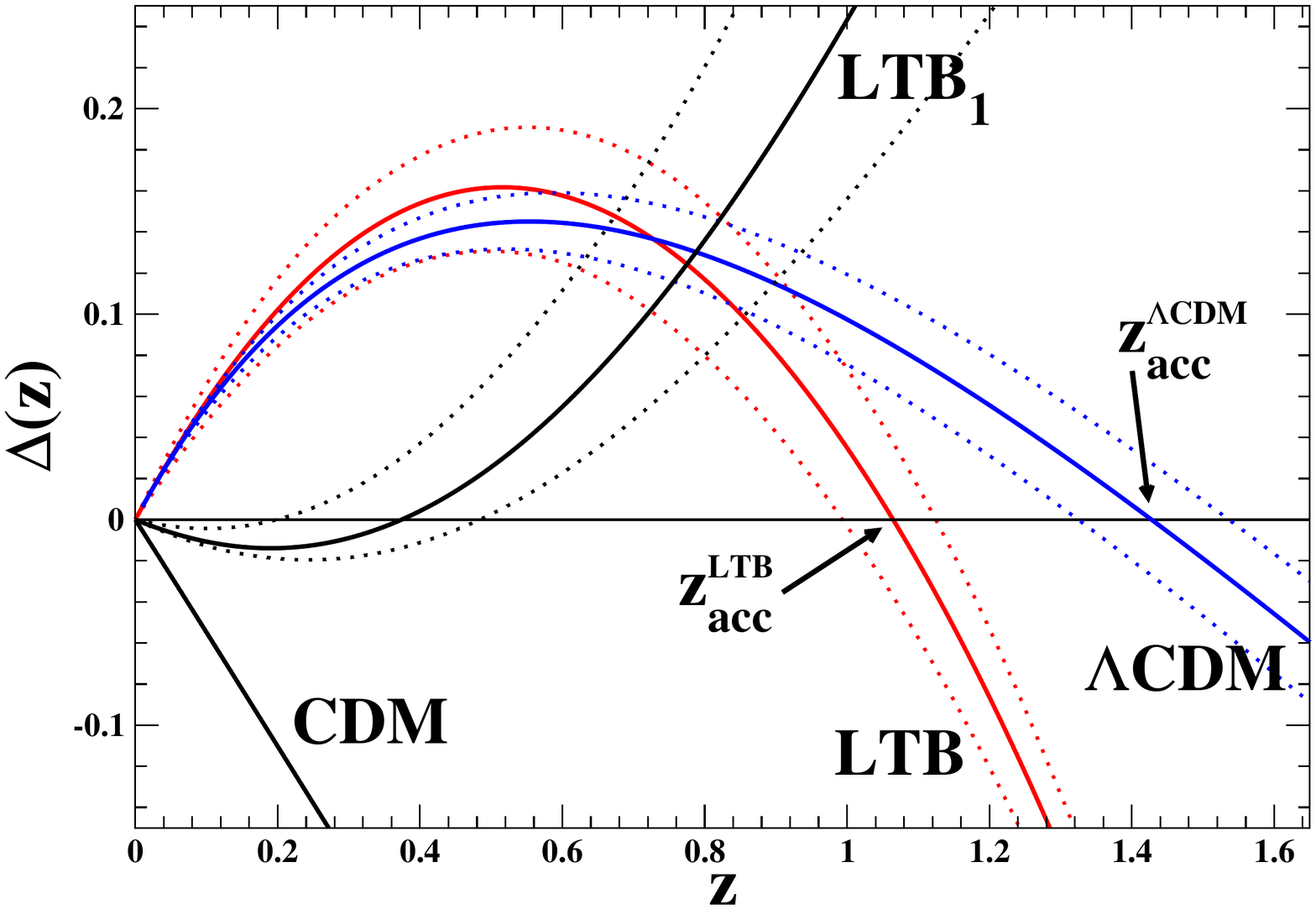}
\caption{The left panel illustrates the  best-fit result for the model characterized by the hyperbolic profile of Eq. (\ref{16}). The corresponding behaviour of $\Da(z)$, for the range of $z$ of interest for this paper, is represented by the curve labelled ${\rm LTB}_1$ reported in the right panel.}
\label{Fig3}
\end{figure}

In order to stress the importance of the parameter $z_{\rm acc}$ let us now consider another possible form of the phenomenological profile $H_0(r)$ appearing in the LTB solution (\ref{4}), for instance the profile\footnote[7]{We thank an anonymous referee for this suggestion. See also \cite{21} for other similar profiles.}
\beq 
H_0(r)= \overline H+ \Da H \tanh \left(r_0-r\over 2 \Da r\right).
\label{16}
\eeq
We can then explicitly check that different models are characterized by largely different values of $z_{\rm acc}$ even within the same class of inhomogenous geometries. By imposing, as before, the phenomenological constraint $H_0(0)= 70 \,{\rm Km \, s}^{-1} {\rm Mpc}^{-1}$ (in order to eliminate $\overline H$), we find that the new profile (\ref{16}) provides indeed a satisfactory three-parameter fit of the Union2 data (see Fig. \ref{Fig3}, left panel), with best fit values $r_0=2500 \pm 322$ Mpc, $\Da r= 2387 \pm 170$ Mpc, $\Da H= 37.5 \pm 2.8 \,{\rm Km \, s}^{-1} {\rm Mpc}^{-1}$, at a confidence level of $95 \%$, with $\chi^2/{\rm d.o.f.}= 1.31$. However, the corresponding value of $z_{\rm acc}$ for this model (called ${\rm LTB}_1$ in Fig. \ref{Fig3})  is significantly different from that of the previous LTB model, and, most important, the behaviour of $\Da(z)$ is exactly the opposite of the standard one, for the range of $z$ of our interest (see Fig. \ref{Fig3}, right panel). We have checked that the value of $\Da (z)$, for ${\rm LTB}_1$, turns back to the standard negative range only for $z \gaq 50$.

%%%%%%%%%%%%%%%%%%%

%%%%%%%%%%%%%%%%%%%%%%%%%%%%

Let us finally comment on the possibility that an off-center position of the observer embedded in a spherically symmetric LTB geometry may significantly affect the determination of $z_{\rm acc}^{\rm LTB}$, thus providing obstructions to a precise discrimination between LTB-based and a more conventional (homogeneous) fit of the SNe data. Indeed, if the observer is located at a distance $r_0 \not=0$ from the center of a spherically symmetry geometry, the corresponding luminosity distance $d_L$ (referred to the position $r_0$) is no longer isotropic but acquires an angular dependence, and this in turn induces an angular dispersion of the value of $z_{\rm acc}$ which depends on $r_0$, and which obviously grows (in modulo) with the growth of $r_0$.

The luminosity distance of a source for off-center observers in a LTB geometry has been computed in \cite{3a} (see also \cite{29a}) as a function of $z$, of the distance $r_0$ from the centre, and of the polar observation angle $\ga$ (referred to $r_0$). We have applied the results of \cite{3a} to compute the directional variation of $z_{\rm acc}$, at fixed values of $r_0$. We have considered, in particular, possible displacements from the centre in the range $r_0 \laq 10^{-2} r_V$, because -- as discussed in \cite{3a} -- higher values of $r_0$ would induce a dipole anisotropy too high to be compatible with present CMB observations. 

\begin{figure}[h!]
\centering
\includegraphics[height=6 cm]{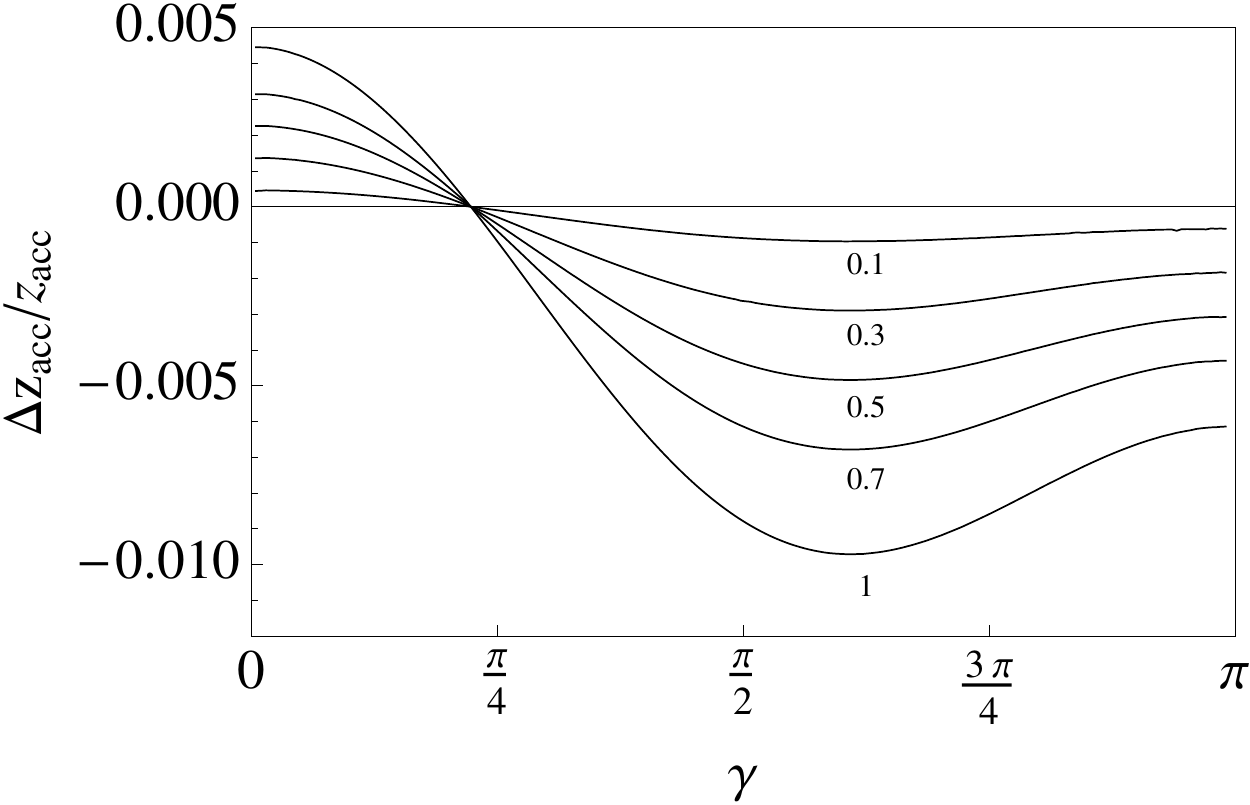}
\caption{The fractional variation of the parameter $z_{\rm acc}$ as a function of the angular direction $\ga$, for different values of the observer's position $r_0$ ranging from $0$ to $10^{-2} r_V$. The numerical labels of the curves are referred to the values of $r_0$, given in units of $10^{-2} r_V$.}
\label{Fig4}
\end{figure}

The results of our exercise are illustrated in Fig. \ref{Fig4}, where we have plotted the fractional variation $ \Da z_{\rm acc}/ z_{\rm acc} \equiv [z_{\rm acc}(r_0, \ga) - z_{\rm acc}(0)]/z_{\rm acc}(0)$, for different values of $r_0$ up to $10^{-2} r_V$, for the LTB model characterized by the parameter $z_{\rm acc}^{\rm LTB}$ of Eq. (\ref{15}). For the normalization of $\mu^{\rm Milne}$ we have consistently used $H_0(r_0)$, but we have checked that using the fixed value $H_0= 70 \,{\rm Km \, s}^{-1} {\rm Mpc}^{-1}$ simply rescales the zero of the difference $ \Da z_{\rm acc}$, without affecting the overall amplitude of the dispersion. As shown in Fig. \ref{Fig4}, the angular variation of $z_{\rm acc}$ induced by $r_0\not=0$ is bounded to be at most at the one-percent level, and has thus a negligible impact on the results of Fig. \ref{Fig2}. 

%%%%%%%%%%%%%%%%%%%

%%%%%%%%%%%%%%%%%%%%%%%%%%%%

In conclusion, we would like to stress again that the inhomogeneous model discussed in this paper should  {\em not} be intended as a realistic alternative to the successful concordance cosmology, but only as a pedagogical example  to learn how to distinguish different fits of SNe data based on different geometrical schemes. To this purpose we have shown, in particular, that in the model of this paper the Universe enters the regime directly affected the accelerated kinematics {\em later} than predicted by the $\La$CDM scenario, i.e. $z_{\rm acc}^{\rm LTB}<z_{\rm acc}^{\La{\rm CDM}}$. Hence, a precise determination of the transition epoch $z_{\rm acc}$ 
(possibly through future extensions of the Hubble diagram to higher values of $z$, or through indirect studies of the transfer function of primordial perturbations \cite{23}), could help us to physically discriminate among statistically equivalent fits.

%%%%%%%%%%%%%%%%%%%%

\section*{Acknowledgements}
One of us (MG) is very grateful to Ido Ben-Dayan, Giovanni  Marozzi, Fabien Nugier and Gabriele Veneziano for many useful discussions on the luminosity distance in the context of inhomogeneous cosmological models.

\end{document}